\documentclass[preprint]{aastex}

\newcommand{\avgm}{\langle M \rangle}

\shorttitle{TIME DELAYS AND QUASAR STRUCTURE}
\shortauthors{MORGAN ET AL.}

\begin{document}

\title{Simultaneous Estimation of Time Delays and Quasar Structure\footnote{B\MakeLowercase{ased on 
observations obtained with the} S\MakeLowercase{mall and} M\MakeLowercase{oderate} A\MakeLowercase{perture} R\MakeLowercase{esearch} T\MakeLowercase{elescope}
S\MakeLowercase{ystem} (SMARTS) 1.3\MakeLowercase{m, which is operated by the} SMARTS C\MakeLowercase{onsortium,
the} A\MakeLowercase{pache} P\MakeLowercase{oint} O\MakeLowercase{bservatory 3.5m telescope, which is 
owned and operated by the} A\MakeLowercase{strophysical} R\MakeLowercase{esearch} C\MakeLowercase{onsortium, the} WIYN O\MakeLowercase{bservatory
which is owned and operated by the} U\MakeLowercase{niversity of} W\MakeLowercase{isconsin}, I\MakeLowercase{ndiana} U\MakeLowercase{niversity}, 
Y\MakeLowercase{ale} U\MakeLowercase{niversity and the} N\MakeLowercase{ational} O\MakeLowercase{ptical} A\MakeLowercase{stronomy} O\MakeLowercase{bservatories} (NOAO), \MakeLowercase{the 
6.5m} M\MakeLowercase{agellan} B\MakeLowercase{aade telescope, which is a collaboration between the observatories 
of the} C\MakeLowercase{arnegie} I\MakeLowercase{nstitution of} W\MakeLowercase{ashington} (OCIW), U\MakeLowercase{niversity of} A\MakeLowercase{rizona}, 
H\MakeLowercase{arvard} U\MakeLowercase{niversity}, U\MakeLowercase{niversity of} M\MakeLowercase{ichigan, and} M\MakeLowercase{assachusetts} I\MakeLowercase{nstitute 
of} T\MakeLowercase{echnology, and observations made with the} NASA/ESA H\MakeLowercase{ubble} S\MakeLowercase{pace} T\MakeLowercase{elescope
for program} HST-GO-9744 \MakeLowercase{of the} S\MakeLowercase{pace} T\MakeLowercase{elescope} S\MakeLowercase{cience} I\MakeLowercase{nstitute,
which is operated by the} A\MakeLowercase{ssociation of} U\MakeLowercase{niversities for} R\MakeLowercase{esearch
in} A\MakeLowercase{stronomy}, I\MakeLowercase{nc., under} NASA \MakeLowercase{contract} NAS 5-26555.
}}

\author{Christopher W. Morgan\altaffilmark{2} and Michael E. Eyler}
\affil{Department of Physics, United States Naval Academy, 572C Holloway Road,
Annapolis, MD 21402}

\author{C.S. Kochanek and Nicholas D. Morgan}
\affil{Department of Astronomy, The Ohio State University, 140 West 18th Avenue, Columbus, OH 43210
-1173}

\author{Emilio E. Falco}
\affil{Harvard-Smithsonian Center for Astrophysics, 60 Garden Street, Cambridge, MA, 02138}

\and

\author{C. Vuissoz, F. Courbin and G. Meylan}
\affil{Laboratoire d'Astrophysique, \'{E}cole Polytechnique F\'{e}d\'{e}rale de Lausanne (EPFL),
Observatoire, 1290 Sauverny, Switzerland}

\altaffiltext{2}{Department of Astronomy, The Ohio State University}

\clearpage

\begin{abstract}
We expand our Bayesian Monte Carlo method for analyzing the light curves of gravitationally
lensed quasars to simultaneously estimate time delays and quasar structure including their
mutual uncertainties.  We apply the method to HE1104--1805 and QJ0158--4325, two doubly-imaged
quasars with microlensing and intrinsic variability on comparable time scales.  For
HE1104--1805 the resulting time delay of $\Delta t_{AB} = t_A - t_B = 162.2_{-5.9}^{+6.3}$~days
and accretion disk size estimate of $\log(r_s / {\rm cm}) = 15.7_{-0.5}^{+0.4}$ at $0.2\mu$m
in the rest frame are consistent with earlier estimates but suggest that existing
methods for estimating time delays in the presence of microlensing underestimate the 
uncertainties.  We are unable to measure a time delay for QJ0158--4325, but the
accretion disk size is $\log(r_s / {\rm cm}) = 14.9\pm0.3$ at $0.3\mu$m in the rest frame.
\end{abstract}


\section{Introduction}

Variability in lensed quasar images comes from two very different sources. Changes in 
the quasar's intrinsic luminosity are observable as correlated variability between 
images, while microlensing by the stars in the lens galaxy produces uncorrelated
variability.  Measurements of the time delays between the lensed images from
the correlated variability can be used to study cosmology 
(e.g. \citealt{Refsdal1964} and recently \citealt{Saha2006,Oguri2007}) 
or the distribution of dark matter in the
lens galaxy \citep[e.g.][]{Kochanek2006,Poindexteretal.2007,Vuissoz2007}.  
The microlensing variability can be used
to study the structure of the quasar, the masses of the stars in the lens
galaxy, and the stellar mass fraction near the lensed images 
\citep{Schechter2002,Wambsganss2006}.  It
is now possible to use microlensing to measure the correlation of accretion
disk size with black hole mass \citep{Morgan2007}, the wavelength dependence of the
size of the accretion disk \citep{Poindexter2007} or the differing sizes
of the thermal and non-thermal X-ray emission regions \citep{Pooley2007,Dai2007}.  

The challenge is that most lensed quasars exhibit both intrinsic and microlensing
variability.  To measure a time delay, one must successfully model and remove the
microlensing variability such that only intrinsic variability remains. If the 
microlensing variability has a sufficiently low amplitude or long timescale, 
it can be ignored \citep[e.g. PG1115-080,][]{Schechter1997}, but this is 
a dangerous assumption for many systems.  \citet{Eigenbrod2005} found that for an $80$~day delay, 
adding microlensing perturbations with an amplitude of 5\% (10\%) to a light curve  
increased the uncertainty in the time delay by a factor of 2 (6).
Existing time delay analyses for lenses with microlensing 
\citep[e.g.][]{Paraficz2006,Kochanek2006,Poindexteretal.2007} depend on the
intrinsic and microlensing variability having different time scales. These analyses also require
that the microlensing variability can be modeled by a simple polynomial function.  This approach
will clearly fail if the two sources of variability have similar time scales 
or if the microlensing variability cannot be easily parameterized.

\begin{deluxetable}{lcccccc}
\tabletypesize{\scriptsize}
\tablecaption{HST Astrometry and Photometry of QJ0158--4325 and HE1104--1805}
\tablehead{Lens & \colhead{Component} 
                 &\multicolumn{2}{c}{Astrometry}
                 &\multicolumn{3}{c}{Photometry}\\
                 \colhead{}
		 &\colhead{}
                 &\colhead{$\Delta\hbox{RA}$}
                 &\colhead{$\Delta\hbox{Dec}$}
                 &\colhead{H=F160W}
                 &\colhead{I=F814W}
                 &\colhead{V=F555W}
                 }
\startdata
QJ0158--4325 & A  &$\equiv 0$ &$\equiv 0$ &$16.47\pm0.03$ &$17.81\pm0.04$ &$18.10\pm0.13$\\
& B  &$-1\farcs156\pm0\farcs003$ &$-0\farcs398\pm0\farcs003$ &$17.27\pm0.03$ &$18.62\pm0.11$ &$18.91\pm0.17$\\
& G  &$-0\farcs780\pm0\farcs016$ &$-0\farcs234\pm0\farcs006$ &$16.67\pm0.13$ &$18.91\pm0.06$ &$20.36\pm0.18$\\
\hline
HE1104--1805 & A  &$\equiv 0$ &$\equiv 0$ &$15.91\pm0.01$ &$16.40\pm0.03$ &$16.92\pm0.06$\\
& B  &$+2\farcs901\pm0\farcs003$ &$-1\farcs332\pm0\farcs003$ &$17.35\pm0.03$ &$17.95\pm0.04$ &$18.70\pm0.08$\\
& G  &$+0\farcs965\pm0\farcs003$ &$-0\farcs500\pm0\farcs003$ &$17.52\pm0.09$ &$20.01\pm0.10$ &$23.26\pm0.27$\\
\enddata
\label{tab:hst}
\end{deluxetable}

In this paper, we present a new technique for simultaneously estimating the time delay and 
structure of lensed quasars that exhibit strong microlensing.  In essence, we assume a 
range of time delays and then determine the likelihood of the implied microlensing
variability using the Bayesian Monte Carlo method of Kochanek (2004, see also \citealt{Kochanek2007}).
This allows us to estimate the time delays and the quasar structural parameters 
simultaneously and include the effects of both phenomena on the parameter
uncertainties.   We apply the method to the two doubly-imaged lenses HE1104--1805 \citep{Wisotzki1993}
and QJ0158--4325 \citep{Morgan1999}.  While HE1104--1805 has a well-measured time
delay \citep{Poindexteretal.2007}, the amplitude of the microlensing
 ($\sim 0.05 \; {\rm mag \; yr^{-1}}$ 
over the past decade) and the fact that it exhibits variability on 
the 6 month scale of the time delay suggest that it is close to
the limit where microlensing polynomial fitting methods \citep{Burud2001,Kochanek2006} 
will break down.  QJ0158--4325 clearly shows both correlated and uncorrelated variability,
but the polynomial methods cannot reliably produce a time delay estimate. 
We describe the data and our models in \S\ref{sec:data}, our new approach in \S\ref{sec:analysis}
and the application to the two systems in \S\ref{sec:results}.  In \S~\ref{sec:discussion}, we 
discuss the results and their limitations.  We assume a flat $\Omega_0=0.3$, $\Lambda_0=0.7$,
$H_0=70$~km~s$^{-1}$~Mpc$^{-1}$ cosmology and that the lens redshift of QJ0158--4325
is $z_l=0.5$.  Reasonable changes in this assumed redshift have negligible consequences for the
results.

\section{Observations and Models}
\label{sec:data}  

We created photometric models for the two systems from the WFPC2 and NICMOS $V$- (F555W), 
$I$- (F814W) and $H$-band (F160W) observations of the two systems by the CfA-Arizona Space 
Telescope Lens Survey (CASTLES) following the methods of \citet{Lehar2000}.  The quasars
are modeled as point sources and each lens galaxy as a de Vaucouleurs profile. We chose 
the de Vaucouleurs profile above other models since it provided the best fit. 
Table~\ref{tab:hst} 
summarizes the fits we use here, where the HE1104--1805 model is updated from that
in \citet{Lehar2000} using a deeper $H$-band image obtained to study the quasar host galaxy
\citep{Yoo2006}.

\begin{figure}
\epsscale{1.0}
\plotone{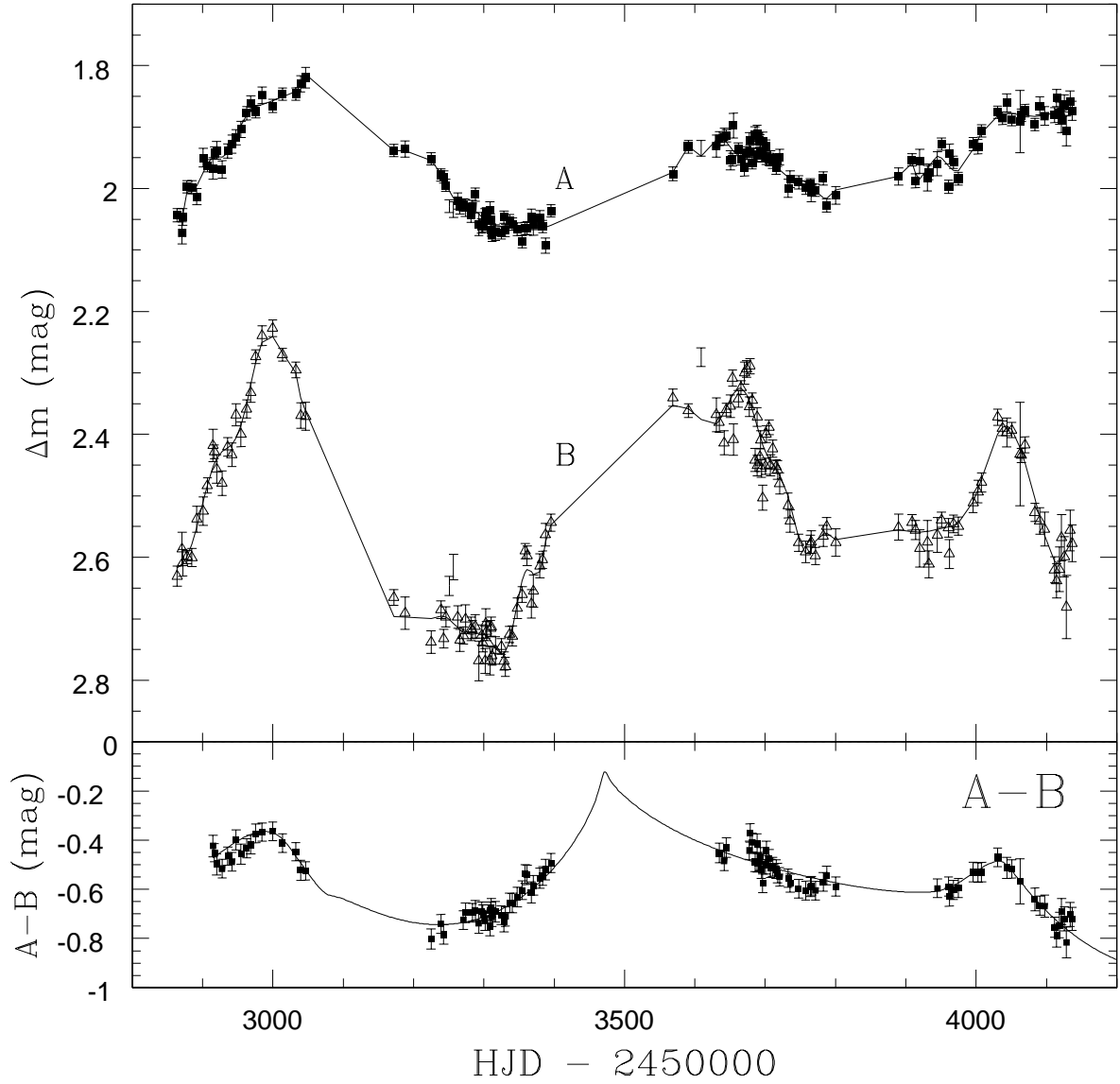}
\caption{Top Panel: QJ0158--4325 $R$-band light curves for images A (squares) 
and B (triangles). The curves are a polynomial fit to guide the
eye. Points with error bars only (no symbols) were not used in the 
analysis. Lower Panel: Example of a good fit to the implied A--B microlensing signal 
for a trial time delay of $\Delta t_{AB} = -20$~days.}
\label{fig:qj0158lightcurve}
\end{figure}

For each system we created a sequence of ten lens models using the {\it lensmodel} 
software package \citep{Keeton2001}.  Each model is the sum of concentric NFW \citep{Navarro1996}
and de Vaucouleurs components, where the NFW component simulates the dark matter halo
and the de Vaucouleurs component represents the galaxy's stellar content.  
We parameterize the model sequence by $f_{M/L}$, the mass of the stellar component
relative to its mass in a constant mass-to-light ratio model with no contribution from
the NFW halo.  We generated model sequences covering the range $0.1 \le f_{M/L} \le 1.0$. 
With their time delay measurement, \citet{Poindexteretal.2007} constrained
the stellar mass fraction of HE1104--1805 to $f_{M/L}=0.30^{+0.04}_{-0.05}$,
but we chose not to apply these limits to $f_{M/L}$ 
for our present calculations.  From these models for the
mass distribution we extract the convergence ($\kappa$), shear ($\gamma$) and stellar 
surface density ($\kappa_*$) for each image and then generate realizations of 
the microlensing magnification patterns at the location of each image using
a variant of the ray-shooting \citep{Schneider1992} method described in \citet{Kochanek2004}.
We assume a stellar mass function of $dN(M)/dM \propto M^{-1.3}$ with a dynamic range 
of 50, which approximates the Galactic stellar mass function of \citet{Gould2000}. 
We present the results either making no assumption about the mean mass $\langle M \rangle$
of the stars or by applying a prior that it lies in the range 
$0.1 \, {\rm M_\sun} \le \avgm  \le 1.0 \, {\rm M_\sun}$.  We used $4096^2$ magnification
patterns with an outer scale of $20 R_E$. We used a prior for the relative motions of the observer,
lens galaxy, lens galaxy stars and the source based on the projection of the CMB
dipole \citep{Kogut1993} for the observer, the stellar velocity dispersion of the lens set by its 
Einstein radius, and rms peculiar velocities for the lens and source of $\sigma_p=235/(1+z)$~km s$^{-1}$
\citep{Kochanek2004}.
We modeled the continuum emission source as a face-on thin accretion disk \citep{Shakura1973}
with the surface brightness profile
\begin{equation}
I(R) \propto \left\{ \exp \left[ \left( R/r_s \right)^{3/4} \right] - 1 \right\}^{-1}. 
\end{equation}
The scale radius $r_s$ is the point where the disk temperature matches the rest frame wavelength 
of our monitoring band, $kT = h c (1+z_s)/\lambda_{obs}$.  For comparisons to other disk models,
the half-light radius of $R_{1/2}=2.44 r_s$ should be used, since \citet{Mortonson2005} have shown
that half-light radii estimated from microlensing depend little on the assumed surface
brightness profile of the disk. 

\begin{figure}
\epsscale{1.0}
\plotone{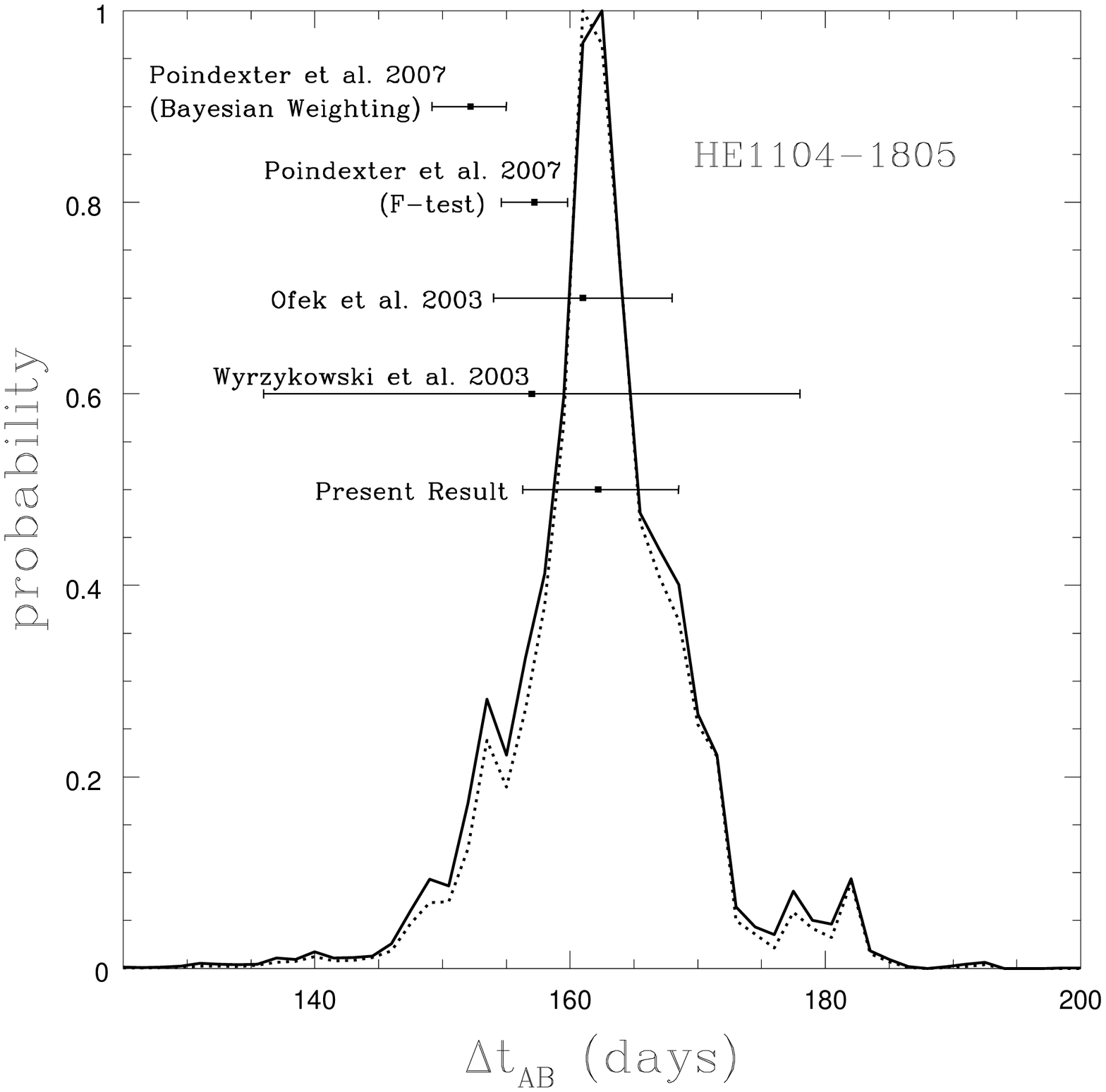}
\caption{Probability distribution for the time delay of HE1104--1805, where
$\Delta t_{AB} = t_A - t_B$.  The solid curve has no prior on the microlens
mass scale while the dotted curve assumes a uniform prior on the mass over
the range $0.1 \, {\rm M_\sun} \le \avgm \le 1.0 \, {\rm M_\sun}$.
Previous measurements of the time delay are also plotted. The dependence
of the \citet{Poindexteretal.2007} results on the statistical test used demonstrates
the limitations of polynomial fitting methods.}
\label{fig:he1104td}
\end{figure}

\begin{figure}
\epsscale{1.0}
\plotone{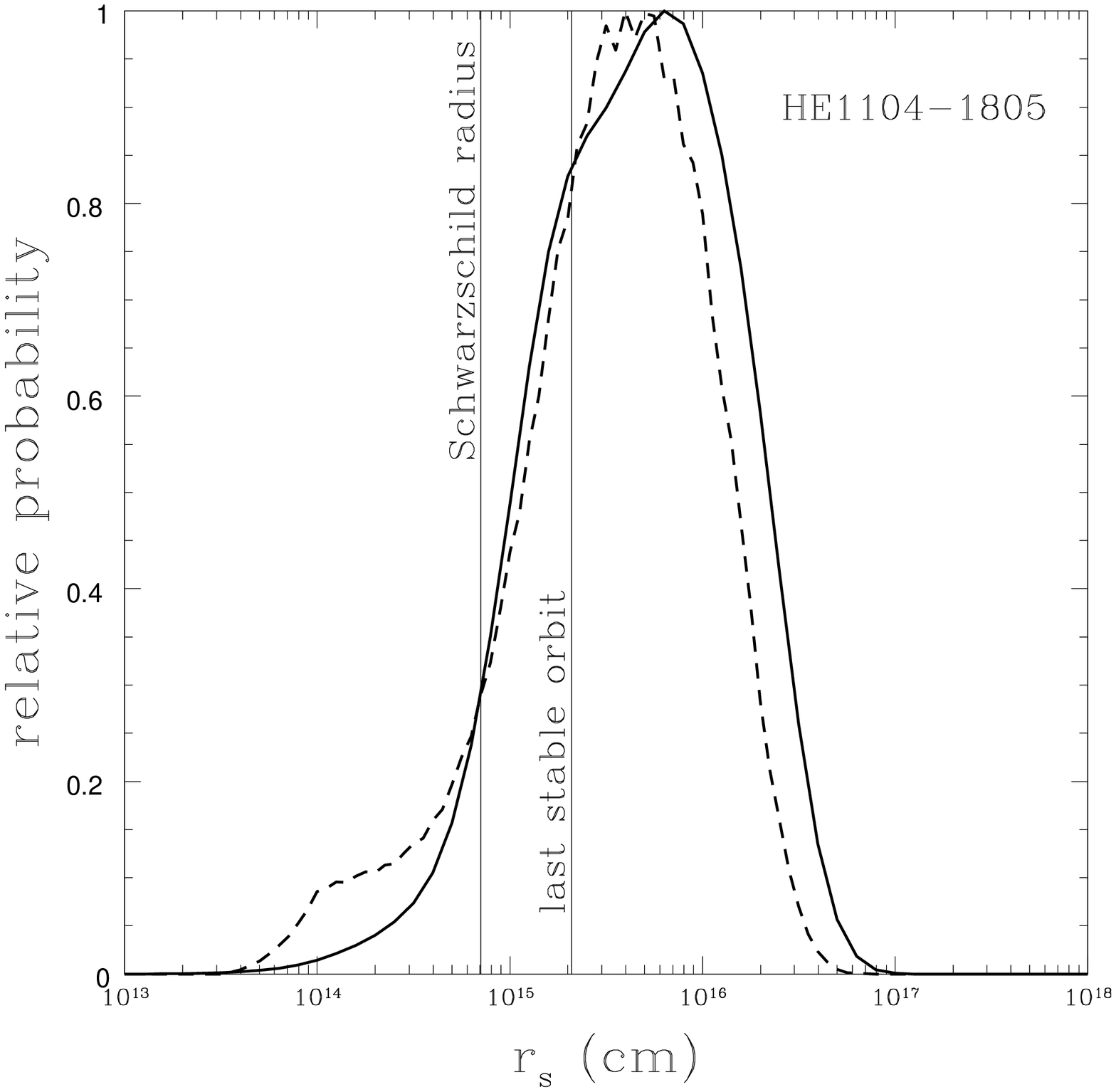}
\caption{Probability distributions for the source size $r_s$ of HE1104--1805.
The dashed curve shows the estimate for $r_s$ with a prior of 
$0.1 \, {\rm M_\sun} \le \avgm \le 1.0 \, {\rm M_\sun}$ on
the mean mass of the microlenses.  The vertical line shows the 
Schwarzschild radius $R_{BH}=2GM_{BH}/c^2$ of the black hole, where
the black hole mass $M_{BH}=2.37 \times 10^9 M_\sun$ was estimated by \citet{Peng2006} 
using the \ion{C}{4} emission line width.  
The last stable orbit for a Schwarzschild black hole is
at $3R_{BH}$.}
\label{fig:he1104rs}
\end{figure}

We monitored QJ0158--4325 in the $R$-band using the SMARTS 1.3m telescope with the  
ANDICAM optical/infrared camera 
\citep{Depoy2003}\footnote{http://www.astronomy.ohio-state.edu/ANDICAM/} and using the 1.2m Euler 
Swiss Telescope as a part of the the COSMOGRAIL\footnote{http://www.cosmograil.org/} project. 
A full description of our monitoring data reduction technique can be found in 
\citet{Kochanek2006}, but we provide a brief summary here.  We model 
the PSF of each quasar image using three nested, elliptical Gaussian components, keeping the 
relative astrometry fixed for all epochs. We use relative photometry, comparing
the flux of each image to the flux of several reference stars in each frame. 
The lens galaxy flux is determined by optimizing its flux in observations with good seeing, 
and the light curves are then measured with the galaxy flux fixed to this optimal value. 
We eliminated all data points taken at seeing conditions worse than $1\farcs7$.  
We dropped 3 epochs that satisfied the seeing conditions but on which the sky was very bright.
On these nights, the flux measurements from the dimmer image B are clearly contaminated by 
sky flux, so we report these measurements but chose not to include them in our microlensing analysis.  
The monitoring data are presented in Table~\ref{tab:qj0158lightcurve}, and the light curves 
are displayed in Figure~\ref{fig:qj0158lightcurve}.  
For HE1104--1805, we use the composite $R$- and 
$V$-band light curve data from \citet{Poindexteretal.2007}.

\section{Joint Monte Carlo Analysis}
\label{sec:analysis}

\begin{figure}
\epsscale{1.0}
\plotone{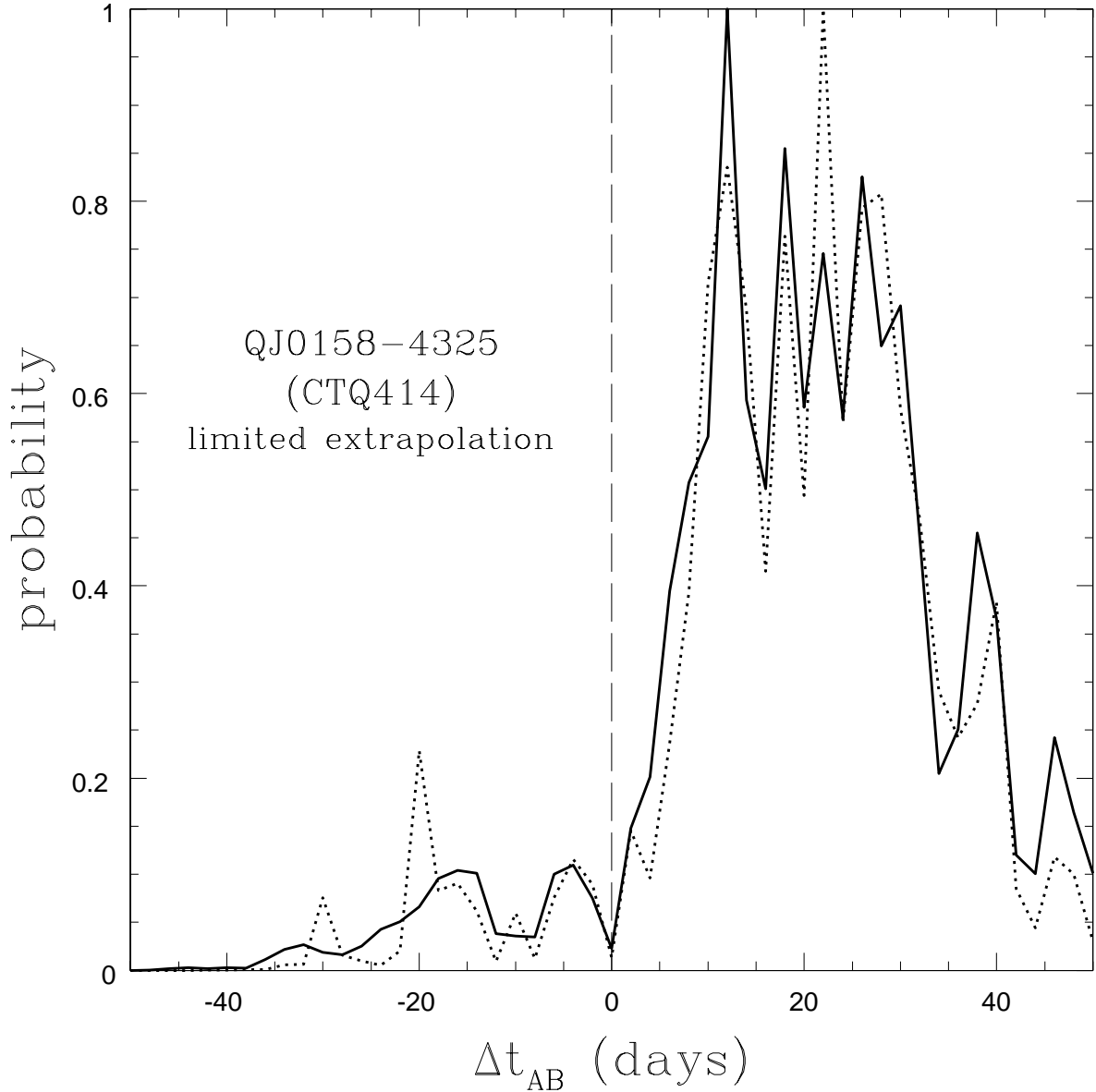}
\caption{Probability distribution for the time delay $\Delta t_{AB} = t_A - t_B$ 
in QJ0158--4325, where we restrict the extrapolation of the light curves to be
less than 7 days.  The solid curve assumes no prior on the microlens mass and the
dashed curve assumes a uniform prior over the range $0.1 \, {\rm M_\sun} \le \avgm \le
1.0 \, {\rm M_\sun}$.  }
\label{fig:qj0158limtd}
\end{figure}

For our analysis we must generate light curve pairs with arbitrary delays.  We
always carry out the shifts on the less variable light curve (image A for both HE1104--1805
and QJ0158--4325).  There are two issues for generating the light curves.  First, 
we must define the algorithms for interpolating and extrapolating a light curve, and
second we must decide how many extrapolated points should be used.
When data points are shifted in the middle of an observing season, we estimate
the flux at the shifted time using linear interpolation between the nearest bracketing
data points.  When shifted points lie in an inter-season gap or beyond either end of the
observed lightcurve, we estimate the flux using extrapolation based on a linear fit to the 
five nearest data points. Data points requiring extrapolation for periods longer than
seven days are discarded. We assign new uncertainties that combine the photometric errors with
the uncertainties due to the temporal distance of the new point from existing data points, 
where we model the source variability using the quasar structure function of \citet{VandenBerk2004}.

For a standard analysis, we wanted to use light curves with the same number of data points
for each time delay and no points extrapolated by more than seven days.
We limit the use of extrapolated points because they lead to a 
delay-dependent change in the statistical weights caused by the steadily 
growing uncertainties from the structure function.   
Given the delay range we wanted to test, we first found the limiting delay
defined by the delay which yielded the minimum number of usable data points given our
seven day extrapolation limit. We then restricted all of the trial light curves to use only the epochs
permissible at the limiting delay. For the 
very long delays of HE1104--1805,
the limiting delay is not the longest delay, because the longest delays shift curves
completely through the inter-season gaps.  
This forced us to restrict the HE1104--1805 analysis to the epochs permitted by both the limiting delay 
and the longest delay,
since some permissible times at the end of the limiting delay lightcurve are beyond 
the extrapolation limit in light curves with longer delays.
We also experimented with simply allowing unlimited extrapolations.  

\begin{figure}
\epsscale{1.0}
\plotone{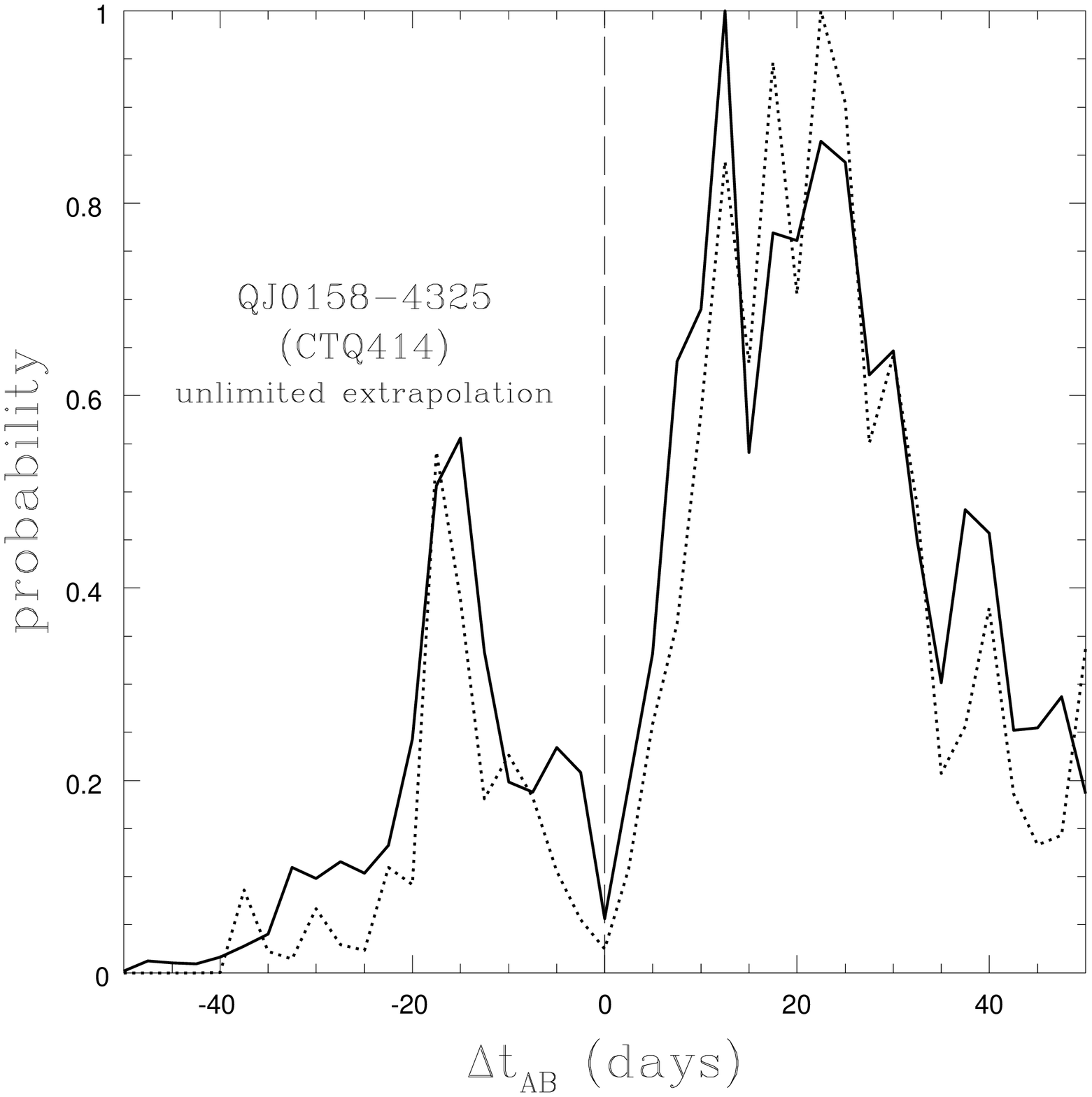}
\caption{Same as Fig.~\ref{fig:qj0158limtd}, but the trial light curves were generated with no limits
on extrapolation.  Note the change in the relative probability of positive and negative delays.}
\label{fig:qj0158td}
\end{figure}

We then analyzed the light curves using the Bayesian Monte Carlo 
method of \citet{Kochanek2004}.  In essence, we randomly select
a time delay $\Delta t$, a lens model and a disk model (size), generate a
microlensing light curve and fit it to the microlensing light
curve implied by the observed data and the selected time delay.
This gives us a $\chi^2$-statistic for the goodness of fit for
the trial $\chi^2(\vec{p},\Delta t)$ given the model parameters
for the microlensing $\vec{p} =$ ($f_{M/L}$, $r_s$, velocities,
masses etc.) and the time delay. Figure~\ref{fig:qj0158lightcurve}
shows an example of a good trial lightcurve fit to QJ 0158--4325
for a delay of $\Delta t_{AB}=-20$~days. In essence, the probability
of time delay $\Delta t$ is the Bayesian integral
\begin{equation}
   P(\Delta t| D) \propto \int P(D| \vec{p}, \Delta t) P(\vec{p}) P(\Delta t) d\vec{p}
\end{equation}
where $P(D|\vec{p}, \Delta t)$ is the probability of fitting the data
in a particular trial, $P(\vec{p})$ sets the priors on the microlensing
variables \citep[see][]{Kochanek2004,Kochanek2007} and $P(\Delta t)$
is the (uniform) prior on the time delay.  The total probability is
then normalized so that $\int P(\Delta t|D)d\Delta t=1$. 
We evaluated the integral
as a Monte Carlo sum over the trial light curves, where we 
created 4 independent sets of magnification patterns for each of the ten macroscopic mass models and 
generated $4 \times 10^6$ trial light curves for each magnification pattern set.

\section{Results}
\label{sec:results}

\citet{Poindexteretal.2007} recently estimated a time delay for HE1104--1805 of 
$\Delta t_{AB} = t_A - t_B = 152.2_{-3.0}^{+2.8}$~days, in the sense that image 
A lags image B, improving on the earlier estimates
by \citet{Ofek2003} and \citet{Wyrzykowski2003}. 
The \citet{Poindexteretal.2007} analysis used the \citet{Kochanek2006} polynomial method
on light curves which combined the published data of \citet{Schechter2003}, 
\citet{Ofek2003} and \citet{Wyrzykowski2003} with new $R$-band monitoring data. In this polynomial method, the source
and microlensing variability are modeled as a set of Legendre polynomials that
are then fit to the light curves.  Ambiguities arise because the value of the
delay depends weakly on the parameterization of the microlensing.  \citet{Poindexteretal.2007} used
a Bayesian weighting scheme for the different polynomial orders, but obtained
$157.2\pm 2.6$~days if they used the F-test to select among the different orders
rather than a Bayesian weighting. The advantage of our present approach is that
it uses a physical model for the microlensing rather than a polynomial parameterization 
of it.

\begin{figure}
\epsscale{1.0}
\plotone{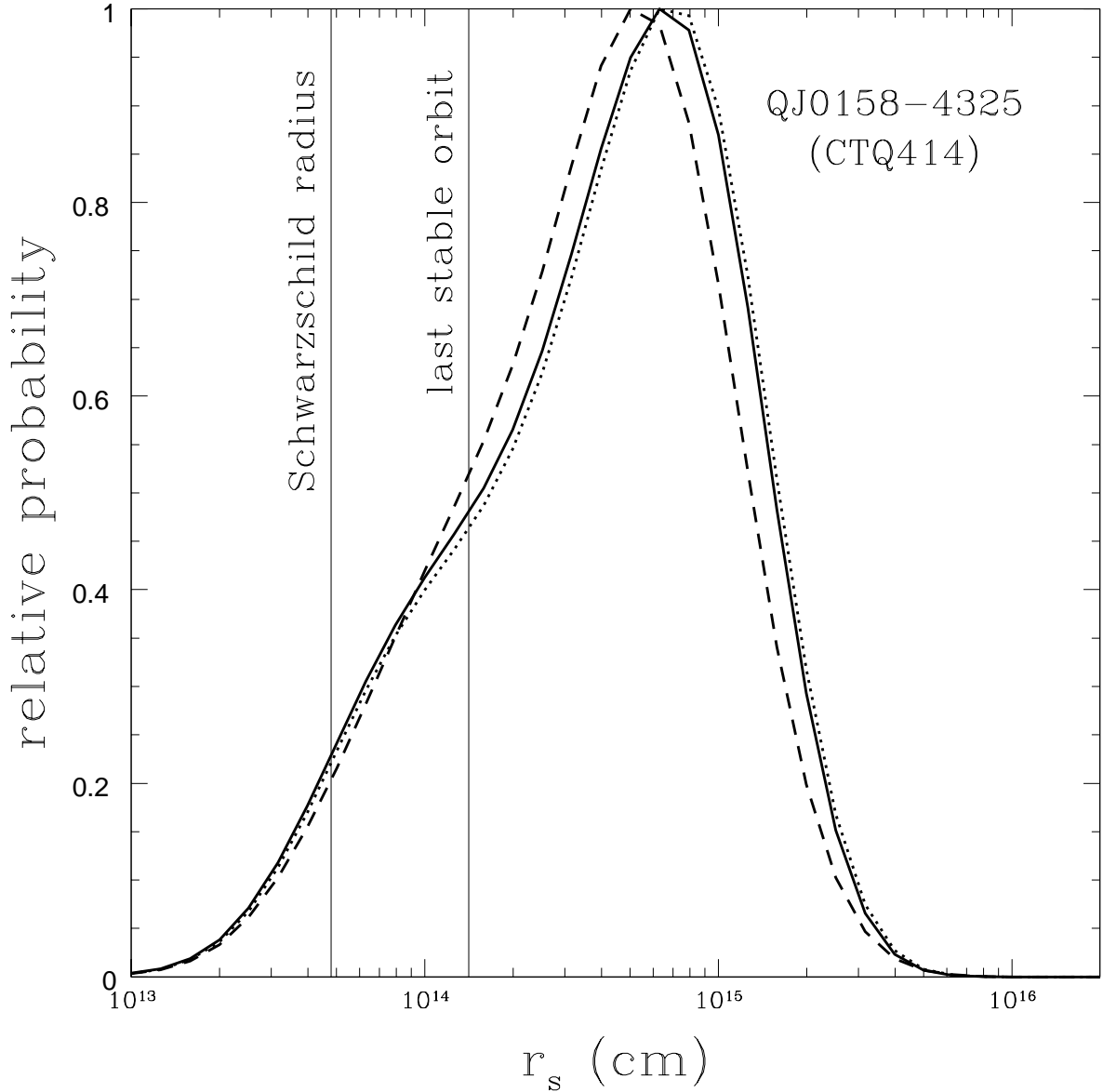}
\caption{Probability distributions for source size $r_s$ of QJ0158--4325.
The dashed (dotted) curve shows the estimate for $r_s$ using the set of negative (positive) 
trial time delays.  The solid curve is the $r_s$ estimate using all trial time delays.
The vertical line shows the 
Schwarzschild radius $R_{BH}=2GM_{BH}/c^2$ of the black hole, where
the black hole mass $M_{BH}=1.6 \times 10^8 M_\sun$ was estimated by \citet{Peng2006} 
using the \ion{Mg}{2} emission line width.   
The last stable orbit for a Schwarzschild black hole is
at $3R_{BH}$.}
\label{fig:qj0158rs}
\end{figure}
  
We applied our joint Monte Carlo analysis technique to HE1104--1805 over a time
interval range of $125 \: {\rm days} \le \Delta t_{AB} \le 200 \: {\rm days}$
with a sampling in 1.5 day intervals and no extrapolation of the light 
curves past 7 days.  Figs.~\ref{fig:he1104td}~and~\ref{fig:he1104rs} show
the resulting probability distributions for the time delay and the disk
size. We find a time delay of 
$\Delta t_{AB} = t_A - t_B = 162.2_{-5.9}^{+6.3}$~days (1$\sigma$)
that is in marginal agreement with the formal \citet{Poindexteretal.2007} result
but in better agreement with the F-test selection of the best polynomial model
than with the Bayesian result.  Assuming $H_0=70 \: {\rm km \: s^{-1} \: Mpc^{-1}}$ and an
inclination angle of $\cos(i)=1/2$, the disk size estimate of 
$\log(r_s / {\rm cm}) = 15.7_{-0.5}^{+0.4}$ at $0.2 \, {\rm \micron}$ in the rest
frame is little changed from
the estimates in \citet{Morgan2007} and \citet{Poindexter2007} which
held the delay fixed to the \citet{Poindexteretal.2007} value.  The 
uncertainties in the size are a factor of $1.6$ larger in the present
analysis.

In models of QJ0158--4325, we expect image A to lead image B by 2--3 weeks depending on
the mass distribution and the actual lens redshift (we assumed $z_l=0.5$).  
For completeness, we tested a full range of 
negative and positive delays $-50 \: {\rm days} \le \Delta t_{AB} \le 50 \: {\rm days}$ 
at 2 day intervals.  We generated the positive and negative delay light curves separately
in order to minimize the number of points lost due to the 7 day extrapolation limit at
the price of making the data used for positive and negative delays somewhat different.
For comparison, we also tried using all points with no extrapolation limits. 

Figs.~\ref{fig:qj0158limtd}~--~\ref{fig:qj0158rs} show the results for the time delays
and the source size.  We have clearly failed to measure a time delay, and
positive delays $\Delta t_{AB} = t_A - t_B$ seem to be favored over negative delays, 
in direct disagreement with the predictions of the lens model.
The relative likelihoods of positive and negative delays depend on the detailed
treatment of the light curves, with the probability of positive delays being lower when 
we use all extrapolated points rather than restricting them to 7 days or less. In this
case, the microlensing simply overwhelms the intrinsic variability.
We expect additional monitoring data to continue to tighten the time delay 
probability distribution, but a successful delay measurement may not be 
possible for many seasons. We succeed, however, in
estimating a size for the quasar despite the uncertainties in the time delay,
finding a size of $\log(r_s / {\rm cm}) = 14.9\pm0.3$ at $0.3 \, {\rm \micron}$
in the rest frame (again, assuming $H_0=70 \: {\rm km \: s^{-1} \: Mpc^{-1}}$ 
and $\cos(i)=1/2$).  The size estimate changes little compared to the uncertainties 
if we limit the analysis to either positive or negative delays. The microlensing 
amplitudes are large enough that interpreting varying amounts of the lower amplitude 
intrinsic variability as microlensing does not change the statistics of the microlensing
enough to significantly affect the size estimate.  
One additional uncertainty in this result is that lens redshift of
QJ0158--4325 is unknown. We experimented with running the Monte Carlo simulation at a
range of lens redshifts ($0.1 \leq z_l \leq 0.9$), 
and we found that the resulting shifts in the $r_s$ estimates 
were negligible relative to the size of the existing uncertainties.  
  
\section{Discussion and Conclusions}
\label{sec:discussion}

\citet{Peng2006} used the width of the \ion{C}{4}~($\lambda 1549$\AA) emission line to estimate the 
black hole mass $M_{BH}=2.37 \times 10^9 {\rm M_\sun}$ in HE1104--1805 and the width of 
\ion{Mg}{2}~($\lambda 2798$\AA) emission line to estimate the black hole mass
$M_{BH}=1.6 \times 10^8 {\rm M_\sun}$ in QJ0158--4325.  Using these black hole masses,
the quasar accretion disk size - black hole mass relation of \citet{Morgan2007} predicts
source sizes at 2500\AA~of $\log(r_s/{\rm cm}) = 15.9 \pm 0.2$ for HE1104--1805 and  
$\log(r_s/{\rm cm}) = 15.2 \pm 0.2$ for QJ0158--4325. If we scale our current
disk size measurements to 2500\AA~using 
the $R_\lambda \propto \lambda^{4/3}$ scaling of thin disk theory and assume an 
inclination angle $\cos i = 1/2$, we find 
$\log(r_s/{\rm cm}) = 15.9^{+0.4}_{-0.5}$ for HE1104--1805 and 
$\log(r_s/{\rm cm}) = 14.8 \pm 0.3$ for QJ0158--4325, fully consistent with the 
predictions of the \citet{Morgan2007} accretion disk size - black hole mass relation.

The mixing of intrinsic and microlensing variability in lensed quasar light
curves can be a serious problem for estimating time delays \citep[e.g.][]{Eigenbrod2005}
and previous microlensing analyses have been restricted to lenses with known time delays.
In HE1104--1805, which must be close to the limits of measuring time delays in
the presence of microlensing, we confirm that the approach of fitting polynomial
models for the microlensing works reasonably well.  However, the dependence of
the delay on the assumed model was a warning sign that the formal errors on
the delays were likely to be underestimates, as was recognized by \citet{Poindexteretal.2007}.
In our new, non-parametric microlensing analysis of HE1104--1805
we find a modestly longer delay of $162.2_{-5.9}^{+6.3}$~days that quantifies
those concerns.   Estimates of the quasar accretion disk size are little
affected by these small shifts in the time delay.
In QJ0158--4325, the microlensing amplitude is larger relative to the intrinsic
variability, and traditional methods for determining delays fail.  Our new
method also fails to measure a delay, but it does allow us to measure the
size of the quasar accretion disk despite the uncertainties in the
time delay. 
  
\acknowledgements
We thank S. Poindexter for helpful discussions about HE1104--1805 and 
D. Will for answers to countless cluster computing questions.  M.E.E. is 
grateful for summer internship support from the USNA Bowman Scholar program.
C.V., F.C. and G.M. acknowledge support from the Swiss National Science
Foundation (SNSF).  
This research made extensive use
of a Beowulf computer cluster obtained through the Cluster Ohio
program of the Ohio Supercomputer Center. Support for program HST-GO-9744 was
provided by NASA through a grant from the Space Telescope Science Institute, which 
is operated by the Association of Universities for Research in Astronomy, Inc., under
NASA contract NAS-5-26666.

{\it Facilities:} \facility{CTIO:2MASS (ANDICAM)}, \facility{HST (NICMOS, ACS)}, 
\facility{Swiss 1.2m Telescope}

\def\hm{\hphantom{-}}
\begin{deluxetable}{cccccc}
\tablecaption{QJ0158--4325 Light curves}
\tablewidth{0pt}
\tablehead{ HJD &\multicolumn{1}{c}{$\chi^2/N_{dof}$}
                &\multicolumn{1}{c}{QSO A (mags)} &\multicolumn{1}{c}{QSO B (mags)} 
                &\multicolumn{1}{c}{$\langle\hbox{Stars}\rangle$}
                &\multicolumn{1}{c}{Source} 
              }
\startdata
$2863.873$ &$  1.17$ &$ 2.043\pm 0.010$ &$ 2.631\pm 0.015$ &$-0.059\pm 0.003$ &SMARTS \\ 
$2870.788$ &$  1.93$ &$ 2.072\pm 0.013$ &$ 2.585\pm 0.020$ &$-0.063\pm 0.003$ &SMARTS \\ 
$2871.813$ &$  0.66$ &$ 2.046\pm 0.014$ &$ 2.609\pm 0.022$ &$-0.063\pm 0.004$ &SMARTS \\ 
$2877.772$ &$  1.01$ &$ 1.997\pm 0.008$ &$ 2.600\pm 0.011$ &$\hm 0.017\pm 0.003$ &SMARTS \\ 
$2884.771$ &$  2.15$ &$ 1.998\pm 0.007$ &$ 2.600\pm 0.010$ &$\hm 0.023\pm 0.003$ &SMARTS \\ 
$2891.770$ &$  0.83$ &$ 2.014\pm 0.013$ &$ 2.537\pm 0.021$ &$-0.068\pm 0.003$ &SMARTS \\ 
$2900.798$ &$  2.26$ &$ 1.950\pm 0.010$ &$ 2.524\pm 0.015$ &$-0.041\pm 0.003$ &SMARTS \\ 
$2906.761$ &$  1.75$ &$ 1.962\pm 0.007$ &$ 2.483\pm 0.009$ &$\hm 0.025\pm 0.003$ &SMARTS \\ 
$2914.653$ &$  0.49$ &$ 1.968\pm 0.018$ &$ 2.417\pm 0.026$ &$-0.045\pm 0.004$ &SMARTS \\ 
$2916.766$ &$  1.00$ &$ 1.942\pm 0.008$ &$ 2.429\pm 0.011$ &$-0.006\pm 0.003$ &SMARTS \\ 
$2919.787$ &$  0.55$ &$ 1.938\pm 0.014$ &$ 2.455\pm 0.023$ &$-0.072\pm 0.003$ &SMARTS \\ 
$2927.729$ &$  2.38$ &$ 1.969\pm 0.009$ &$ 2.479\pm 0.013$ &$-0.022\pm 0.003$ &SMARTS \\ 
$2935.680$ &$  2.25$ &$ 1.939\pm 0.008$ &$ 2.420\pm 0.010$ &$\hm 0.010\pm 0.003$ &SMARTS \\ 
$2941.674$ &$  3.82$ &$ 1.928\pm 0.008$ &$ 2.433\pm 0.010$ &$\hm 0.018\pm 0.003$ &SMARTS \\ 
$2947.635$ &$  0.73$ &$ 1.916\pm 0.012$ &$ 2.368\pm 0.018$ &$-0.054\pm 0.003$ &SMARTS \\ 
$2954.626$ &$  0.85$ &$ 1.903\pm 0.013$ &$ 2.400\pm 0.020$ &$-0.064\pm 0.004$ &SMARTS \\ 
$2962.598$ &$  1.28$ &$ 1.878\pm 0.010$ &$ 2.359\pm 0.013$ &$-0.038\pm 0.003$ &SMARTS \\ 
$2968.621$ &$  3.01$ &$ 1.862\pm 0.007$ &$ 2.332\pm 0.009$ &$-0.008\pm 0.003$ &SMARTS \\ 
$2975.577$ &$  2.02$ &$ 1.875\pm 0.007$ &$ 2.273\pm 0.008$ &$\hm 0.012\pm 0.003$ &SMARTS \\ 
$2984.541$ &$  3.28$ &$ 1.848\pm 0.007$ &$ 2.239\pm 0.009$ &$\hm 0.016\pm 0.003$ &SMARTS \\ 
$2999.619$ &$  1.54$ &$ 1.866\pm 0.009$ &$ 2.227\pm 0.011$ &$\hm 0.010\pm 0.003$ &SMARTS \\ 
$3013.587$ &$  0.92$ &$ 1.846\pm 0.008$ &$ 2.271\pm 0.011$ &$-0.030\pm 0.003$ &SMARTS \\ 
$3032.567$ &$  1.70$ &$ 1.846\pm 0.008$ &$ 2.295\pm 0.010$ &$\hm 0.011\pm 0.003$ &SMARTS \\ 
$3039.534$ &$  1.34$ &$ 1.830\pm 0.011$ &$ 2.369\pm 0.018$ &$-0.068\pm 0.003$ &SMARTS \\ 
$3046.543$ &$  4.29$ &$ 1.820\pm 0.008$ &$ 2.371\pm 0.011$ &$\hm 0.006\pm 0.003$ &SMARTS \\ 
$3171.930$ &$  0.92$ &$ 1.938\pm 0.008$ &$ 2.666\pm 0.013$ &$-0.010\pm 0.003$ &SMARTS \\ 
$3187.869$ &$  0.65$ &$ 1.934\pm 0.013$ &$ 2.694\pm 0.026$ &$-0.068\pm 0.003$ &SMARTS \\ 
$3224.869$ &$  0.70$ &$ 1.950\pm 0.010$ &$ 2.743\pm 0.018$ &$-0.024\pm 0.003$ &SMARTS \\ 
$3238.825$ &$  0.99$ &$ 1.977\pm 0.009$ &$ 2.688\pm 0.014$ &$\hm 0.004\pm 0.003$ &SMARTS \\ 
$3242.777$ &$  0.76$ &$ 1.981\pm 0.009$ &$ 2.732\pm 0.015$ &$-0.026\pm 0.003$ &SMARTS \\ 
$3245.784$ &$  0.61$ &$ 1.994\pm 0.010$ &$ 2.698\pm 0.016$ &$-0.051\pm 0.003$ &SMARTS \\ 
$3250.779$ &$  1.29$ &$( 2.029\pm 0.009)$ &$( 2.646\pm 0.014)$ &$-0.042\pm 0.003$ &SMARTS \\ 
$3256.852$ &$  0.80$ &$( 2.035\pm 0.013)$ &$( 2.616\pm 0.021)$ &$-0.044\pm 0.003$ &SMARTS \\ 
$3262.779$ &$  3.42$ &$ 2.020\pm 0.007$ &$ 2.697\pm 0.010$ &$\hm 0.020\pm 0.003$ &SMARTS \\ 
$3265.790$ &$  1.82$ &$ 2.030\pm 0.008$ &$ 2.735\pm 0.013$ &$-0.002\pm 0.003$ &SMARTS \\ 
$3270.792$ &$  1.46$ &$ 2.027\pm 0.008$ &$ 2.727\pm 0.011$ &$\hm 0.013\pm 0.003$ &SMARTS \\ 
$3273.731$ &$  1.68$ &$ 2.031\pm 0.011$ &$ 2.701\pm 0.018$ &$-0.048\pm 0.003$ &SMARTS \\ 
$3281.769$ &$  1.52$ &$ 2.043\pm 0.010$ &$ 2.716\pm 0.016$ &$-0.037\pm 0.003$ &SMARTS \\ 
$3283.759$ &$  1.65$ &$ 2.029\pm 0.008$ &$ 2.718\pm 0.011$ &$\hm 0.001\pm 0.003$ &SMARTS \\ 
$3287.633$ &$  0.73$ &$ 2.009\pm 0.011$ &$ 2.715\pm 0.019$ &$-0.015\pm 0.003$ &SMARTS \\ 
$3292.735$ &$  2.96$ &$ 2.057\pm 0.011$ &$ 2.769\pm 0.019$ &$-0.038\pm 0.003$ &SMARTS \\ 
$3296.721$ &$  2.25$ &$ 2.061\pm 0.008$ &$ 2.726\pm 0.011$ &$\hm 0.009\pm 0.003$ &SMARTS \\ 
$3298.691$ &$  1.36$ &$ 2.060\pm 0.008$ &$ 2.740\pm 0.012$ &$-0.001\pm 0.003$ &SMARTS \\ 
$3301.686$ &$  0.84$ &$ 2.044\pm 0.011$ &$ 2.769\pm 0.020$ &$-0.058\pm 0.003$ &SMARTS \\ 
$3302.791$ &$  0.98$ &$ 2.052\pm 0.013$ &$ 2.707\pm 0.024$ &$-0.016\pm 0.003$ &EULER \\ 
$3303.690$ &$  0.80$ &$ 2.041\pm 0.010$ &$ 2.720\pm 0.016$ &$-0.052\pm 0.003$ &SMARTS \\ 
$3308.686$ &$  2.29$ &$ 2.035\pm 0.009$ &$ 2.768\pm 0.015$ &$-0.016\pm 0.003$ &EULER \\ 
$3309.642$ &$  0.58$ &$ 2.051\pm 0.009$ &$ 2.717\pm 0.013$ &$-0.026\pm 0.003$ &SMARTS \\ 
$3310.580$ &$  1.51$ &$ 2.067\pm 0.008$ &$ 2.713\pm 0.012$ &$\hm 0.025\pm 0.003$ &EULER \\ 
$3311.639$ &$  1.88$ &$ 2.076\pm 0.008$ &$ 2.760\pm 0.011$ &$\hm 0.014\pm 0.003$ &SMARTS \\ 
$3316.694$ &$  0.59$ &$ 2.071\pm 0.013$ &$ 2.752\pm 0.022$ &$-0.050\pm 0.003$ &SMARTS \\ 
$3324.630$ &$  1.37$ &$ 2.072\pm 0.008$ &$ 2.745\pm 0.011$ &$\hm 0.005\pm 0.003$ &SMARTS \\ 
$3328.616$ &$  0.66$ &$ 2.046\pm 0.009$ &$ 2.769\pm 0.015$ &$-0.018\pm 0.003$ &SMARTS \\ 
$3330.629$ &$  1.05$ &$ 2.068\pm 0.009$ &$ 2.779\pm 0.015$ &$-0.034\pm 0.003$ &SMARTS \\ 
$3336.600$ &$  0.88$ &$ 2.052\pm 0.009$ &$ 2.726\pm 0.013$ &$-0.035\pm 0.003$ &SMARTS \\ 
$3340.603$ &$  0.96$ &$ 2.058\pm 0.010$ &$ 2.728\pm 0.016$ &$-0.008\pm 0.003$ &SMARTS \\ 
$3347.574$ &$  0.83$ &$ 2.065\pm 0.010$ &$ 2.685\pm 0.017$ &$-0.010\pm 0.003$ &SMARTS \\ 
$3354.564$ &$  1.45$ &$ 2.086\pm 0.008$ &$ 2.661\pm 0.011$ &$\hm 0.007\pm 0.003$ &SMARTS \\ 
$3358.560$ &$  1.00$ &$ 2.064\pm 0.009$ &$ 2.592\pm 0.012$ &$-0.012\pm 0.003$ &SMARTS \\ 
$3361.553$ &$  0.73$ &$ 2.063\pm 0.011$ &$ 2.600\pm 0.016$ &$-0.043\pm 0.003$ &SMARTS \\ 
$3367.572$ &$  0.62$ &$ 2.046\pm 0.014$ &$ 2.677\pm 0.023$ &$-0.066\pm 0.004$ &SMARTS \\ 
$3370.582$ &$  2.26$ &$ 2.057\pm 0.011$ &$ 2.658\pm 0.018$ &$-0.030\pm 0.003$ &SMARTS \\ 
$3379.581$ &$  0.55$ &$ 2.046\pm 0.012$ &$ 2.619\pm 0.020$ &$-0.017\pm 0.003$ &SMARTS \\ 
$3383.581$ &$  1.32$ &$ 2.062\pm 0.009$ &$ 2.604\pm 0.012$ &$-0.002\pm 0.003$ &SMARTS \\ 
$3387.563$ &$  0.63$ &$ 2.092\pm 0.012$ &$ 2.566\pm 0.018$ &$-0.043\pm 0.003$ &SMARTS \\ 
$3395.547$ &$  0.86$ &$ 2.036\pm 0.010$ &$ 2.544\pm 0.014$ &$-0.062\pm 0.003$ &SMARTS \\ 
$3568.889$ &$  1.66$ &$ 1.976\pm 0.009$ &$ 2.340\pm 0.011$ &$-0.018\pm 0.003$ &SMARTS \\ 
$3590.890$ &$  1.19$ &$ 1.932\pm 0.008$ &$ 2.362\pm 0.010$ &$-0.003\pm 0.003$ &SMARTS \\ 
$3608.820$ &$  4.55$ &$( 1.935\pm 0.006)$ &$( 2.274\pm 0.007)$ &$\hm 0.112\pm 0.002$ &EULER \\ 
$3630.626$ &$  0.75$ &$ 1.929\pm 0.018$ &$ 2.370\pm 0.027$ &$-0.053\pm 0.004$ &SMARTS \\ 
$3634.802$ &$  0.86$ &$ 1.918\pm 0.011$ &$ 2.382\pm 0.016$ &$-0.063\pm 0.003$ &SMARTS \\ 
$3641.753$ &$  2.01$ &$ 1.916\pm 0.010$ &$ 2.414\pm 0.014$ &$-0.016\pm 0.003$ &SMARTS \\ 
$3644.792$ &$  0.91$ &$ 1.914\pm 0.008$ &$ 2.360\pm 0.010$ &$-0.007\pm 0.003$ &SMARTS \\ 
$3653.595$ &$  1.43$ &$ 1.953\pm 0.009$ &$ 2.309\pm 0.011$ &$-0.028\pm 0.003$ &SMARTS \\ 
$3661.717$ &$  0.77$ &$ 1.936\pm 0.010$ &$ 2.342\pm 0.013$ &$-0.067\pm 0.003$ &SMARTS \\ 
$3665.707$ &$  0.91$ &$ 1.952\pm 0.009$ &$ 2.324\pm 0.011$ &$-0.041\pm 0.003$ &SMARTS \\ 
$3670.620$ &$  1.53$ &$ 1.964\pm 0.011$ &$ 2.302\pm 0.015$ &$-0.022\pm 0.003$ &SMARTS \\ 
$3673.595$ &$  0.81$ &$ 1.942\pm 0.011$ &$ 2.294\pm 0.013$ &$-0.037\pm 0.003$ &SMARTS \\ 
$3677.661$ &$  3.41$ &$ 1.922\pm 0.007$ &$ 2.355\pm 0.009$ &$\hm 0.131\pm 0.002$ &EULER \\ 
$3678.644$ &$  1.28$ &$ 1.939\pm 0.009$ &$ 2.290\pm 0.011$ &$-0.004\pm 0.003$ &SMARTS \\ 
$3681.643$ &$  0.88$ &$ 1.958\pm 0.009$ &$ 2.345\pm 0.012$ &$-0.033\pm 0.003$ &SMARTS \\ 
$3685.669$ &$  5.14$ &$ 1.914\pm 0.006$ &$ 2.442\pm 0.008$ &$\hm 0.079\pm 0.003$ &EULER \\ 
$3688.633$ &$  0.50$ &$ 1.944\pm 0.011$ &$ 2.373\pm 0.015$ &$-0.064\pm 0.003$ &SMARTS \\ 
$3688.676$ &$  4.26$ &$ 1.910\pm 0.006$ &$ 2.451\pm 0.008$ &$\hm 0.069\pm 0.003$ &EULER \\ 
$3692.631$ &$  2.72$ &$ 1.923\pm 0.006$ &$ 2.436\pm 0.008$ &$\hm 0.094\pm 0.003$ &EULER \\ 
$3693.560$ &$  2.10$ &$ 1.941\pm 0.008$ &$ 2.409\pm 0.010$ &$-0.012\pm 0.003$ &SMARTS \\ 
$3694.631$ &$  2.43$ &$ 1.946\pm 0.007$ &$ 2.455\pm 0.009$ &$\hm 0.113\pm 0.002$ &EULER \\ 
$3696.623$ &$  5.99$ &$ 1.923\pm 0.006$ &$ 2.503\pm 0.008$ &$\hm 0.135\pm 0.002$ &EULER \\ 
$3700.648$ &$  2.38$ &$ 1.931\pm 0.006$ &$ 2.452\pm 0.009$ &$\hm 0.122\pm 0.002$ &EULER \\ 
$3701.605$ &$  1.01$ &$ 1.951\pm 0.010$ &$ 2.400\pm 0.013$ &$-0.021\pm 0.003$ &SMARTS \\ 
$3705.652$ &$  1.12$ &$ 1.952\pm 0.009$ &$ 2.389\pm 0.011$ &$-0.004\pm 0.003$ &SMARTS \\ 
$3707.706$ &$  2.82$ &$ 1.951\pm 0.007$ &$ 2.450\pm 0.010$ &$\hm 0.122\pm 0.002$ &EULER \\ 
$3710.605$ &$  0.99$ &$ 1.957\pm 0.010$ &$ 2.424\pm 0.014$ &$-0.012\pm 0.003$ &SMARTS \\ 
$3715.693$ &$  2.21$ &$ 1.967\pm 0.006$ &$ 2.452\pm 0.007$ &$\hm 0.105\pm 0.002$ &EULER \\ 
$3717.573$ &$  0.68$ &$ 1.953\pm 0.010$ &$ 2.459\pm 0.015$ &$-0.061\pm 0.003$ &SMARTS \\ 
$3720.644$ &$  3.80$ &$ 1.949\pm 0.007$ &$ 2.480\pm 0.009$ &$\hm 0.062\pm 0.003$ &EULER \\ 
$3732.631$ &$  4.36$ &$ 1.999\pm 0.007$ &$ 2.516\pm 0.010$ &$\hm 0.121\pm 0.002$ &EULER \\ 
$3735.616$ &$  4.39$ &$ 1.985\pm 0.006$ &$ 2.541\pm 0.009$ &$\hm 0.126\pm 0.002$ &EULER \\ 
$3747.614$ &$  3.03$ &$ 1.990\pm 0.006$ &$ 2.576\pm 0.008$ &$\hm 0.079\pm 0.003$ &EULER \\ 
$3757.606$ &$  1.92$ &$ 1.998\pm 0.008$ &$ 2.591\pm 0.013$ &$\hm 0.088\pm 0.003$ &EULER \\ 
$3764.534$ &$  0.80$ &$ 1.994\pm 0.010$ &$ 2.579\pm 0.015$ &$-0.033\pm 0.003$ &SMARTS \\ 
$3765.570$ &$  6.22$ &$ 2.007\pm 0.006$ &$ 2.575\pm 0.007$ &$\hm 0.130\pm 0.002$ &EULER \\ 
$3771.604$ &$  2.31$ &$ 2.003\pm 0.006$ &$ 2.598\pm 0.009$ &$\hm 0.091\pm 0.003$ &EULER \\ 
$3782.557$ &$  1.59$ &$ 1.984\pm 0.008$ &$ 2.565\pm 0.013$ &$\hm 0.032\pm 0.003$ &EULER \\ 
$3787.531$ &$  2.82$ &$ 2.027\pm 0.006$ &$ 2.549\pm 0.009$ &$\hm 0.098\pm 0.003$ &EULER \\ 
$3800.519$ &$  2.23$ &$ 2.011\pm 0.010$ &$ 2.576\pm 0.015$ &$\hm 0.032\pm 0.003$ &EULER \\ 
$3889.922$ &$  2.68$ &$ 1.980\pm 0.008$ &$ 2.551\pm 0.013$ &$\hm 0.097\pm 0.003$ &EULER \\ 
$3908.910$ &$  2.44$ &$ 1.953\pm 0.006$ &$ 2.543\pm 0.008$ &$\hm 0.128\pm 0.002$ &EULER \\ 
$3913.839$ &$  2.94$ &$ 1.987\pm 0.007$ &$ 2.555\pm 0.009$ &$\hm 0.102\pm 0.003$ &EULER \\ 
$3919.880$ &$  5.01$ &$ 1.956\pm 0.009$ &$ 2.585\pm 0.014$ &$\hm 0.029\pm 0.003$ &EULER \\ 
$3930.856$ &$  0.49$ &$ 1.983\pm 0.021$ &$ 2.577\pm 0.036$ &$-0.016\pm 0.004$ &SMARTS \\ 
$3932.920$ &$  1.40$ &$ 1.973\pm 0.011$ &$ 2.611\pm 0.019$ &$\hm 0.007\pm 0.003$ &EULER \\ 
$3950.832$ &$  2.46$ &$ 1.928\pm 0.006$ &$ 2.539\pm 0.008$ &$\hm 0.140\pm 0.002$ &EULER \\ 
$3960.817$ &$  1.23$ &$ 1.997\pm 0.010$ &$ 2.552\pm 0.014$ &$-0.051\pm 0.003$ &SMARTS \\ 
$3961.922$ &$  6.26$ &$ 1.944\pm 0.006$ &$ 2.594\pm 0.009$ &$\hm 0.093\pm 0.003$ &EULER \\ 
$3967.838$ &$  0.85$ &$ 1.957\pm 0.010$ &$ 2.546\pm 0.014$ &$-0.026\pm 0.003$ &SMARTS \\ 
$3974.783$ &$  1.11$ &$ 1.984\pm 0.010$ &$ 2.550\pm 0.014$ &$-0.030\pm 0.003$ &SMARTS \\ 
$3995.765$ &$  0.88$ &$ 1.927\pm 0.011$ &$ 2.511\pm 0.016$ &$-0.034\pm 0.003$ &SMARTS \\ 
$4002.702$ &$  1.03$ &$ 1.931\pm 0.012$ &$ 2.494\pm 0.018$ &$-0.043\pm 0.003$ &SMARTS \\ 
$4007.696$ &$  0.70$ &$ 1.905\pm 0.010$ &$ 2.479\pm 0.015$ &$-0.032\pm 0.003$ &SMARTS \\ 
$4030.632$ &$  0.82$ &$ 1.876\pm 0.010$ &$ 2.372\pm 0.013$ &$-0.020\pm 0.003$ &SMARTS \\ 
$4037.600$ &$  0.92$ &$ 1.886\pm 0.009$ &$ 2.390\pm 0.012$ &$-0.025\pm 0.003$ &SMARTS \\ 
$4043.600$ &$  0.48$ &$ 1.861\pm 0.015$ &$ 2.397\pm 0.023$ &$-0.063\pm 0.004$ &SMARTS \\ 
$4050.620$ &$  1.34$ &$ 1.888\pm 0.008$ &$ 2.393\pm 0.011$ &$-0.022\pm 0.003$ &SMARTS \\ 
$4062.536$ &$  0.47$ &$ 1.891\pm 0.051$ &$ 2.432\pm 0.084$ &$\hm 0.109\pm 0.005$ &SMARTS \\ 
$4064.646$ &$  0.84$ &$ 1.881\pm 0.009$ &$ 2.434\pm 0.012$ &$-0.011\pm 0.003$ &SMARTS \\ 
$4069.580$ &$  0.82$ &$ 1.872\pm 0.009$ &$ 2.419\pm 0.013$ &$-0.049\pm 0.003$ &SMARTS \\ 
$4083.542$ &$  0.69$ &$ 1.895\pm 0.010$ &$ 2.527\pm 0.015$ &$-0.037\pm 0.003$ &SMARTS \\ 
$4090.621$ &$  2.29$ &$ 1.865\pm 0.010$ &$ 2.541\pm 0.014$ &$-0.037\pm 0.003$ &SMARTS \\ 
$4097.556$ &$  0.40$ &$ 1.882\pm 0.015$ &$ 2.554\pm 0.027$ &$-0.059\pm 0.004$ &SMARTS \\ 
$4111.601$ &$  1.23$ &$ 1.880\pm 0.011$ &$ 2.622\pm 0.019$ &$-0.033\pm 0.003$ &SMARTS \\ 
$4114.561$ &$  1.08$ &$ 1.842\pm 0.013$ &$ 2.664\pm 0.027$ &$-0.037\pm 0.003$ &SMARTS \\ 
$4118.593$ &$  0.48$ &$ 1.873\pm 0.018$ &$ 2.628\pm 0.036$ &$-0.055\pm 0.004$ &SMARTS \\ 
$4121.529$ &$  0.57$ &$ 1.888\pm 0.020$ &$ 2.571\pm 0.038$ &$-0.049\pm 0.004$ &SMARTS \\ 
$4125.526$ &$  0.66$ &$ 1.863\pm 0.017$ &$ 2.602\pm 0.033$ &$-0.047\pm 0.004$ &SMARTS \\ 
$4128.574$ &$  0.51$ &$ 1.900\pm 0.024$ &$ 2.696\pm 0.052$ &$-0.047\pm 0.004$ &SMARTS \\ 
$4133.581$ &$  0.45$ &$ 1.858\pm 0.017$ &$ 2.557\pm 0.033$ &$-0.051\pm 0.004$ &SMARTS \\ 
$4136.572$ &$  0.49$ &$ 1.873\pm 0.015$ &$ 2.578\pm 0.030$ &$-0.068\pm 0.004$ &SMARTS \\ 
$4301.869$ &$  1.05$ &$ 1.815\pm 0.008$ &$ 2.565\pm 0.012$ &$\hm 0.006\pm 0.003$ &SMARTS \\ 
$4307.824$ &$  0.70$ &$ 1.794\pm 0.009$ &$ 2.571\pm 0.016$ &$-0.041\pm 0.003$ &SMARTS \\ 

\enddata
\tablecomments{HJD is the Heliocentric Julian Day -- 2450000 days.
The goodness of fit of the image, $\chi^2/N_{dof}$, is used to rescale the
formal uncertainties when greater than unity.   The QSO A\&B
columns give the magnitudes of the quasar images relative to the
comparison stars. The $\langle\hbox{Stars}\rangle$ column gives the
mean magnitude of the standard stars for that epoch relative to
their mean for all epochs. A few points in the lightcurves
(in parentheses) were not used in the analysis. }
\label{tab:qj0158lightcurve}
\end{deluxetable}

\end{document}